\documentclass{article}

\usepackage{microtype}
\usepackage{graphicx}
\usepackage{subfigure}
\usepackage{booktabs} 
\usepackage{enumitem}

\usepackage{comment}

\usepackage{hyperref}

\usepackage[accepted]{icml2024}

\usepackage{amsmath}
\usepackage{amssymb}
\usepackage{mathtools}
\usepackage{amsthm}

\usepackage[capitalize,noabbrev]{cleveref}

\theoremstyle{plain}

\theoremstyle{definition}

\theoremstyle{remark}

\usepackage[textsize=tiny]{todonotes}

\icmltitlerunning{LLM Multi-Agent Systems: Challenges and Open Problems
}

\begin{document}


\newcommand{\switch}[1]{%
   \ifthenelse{\equal{#1}{0}}{\renewcommand{\review}[1]{}}{}
   \ifthenelse{\equal{#1}{0}}{\renewcommand{\resolve}[1]{}}{}}
\switch{1}


\twocolumn[
\icmltitle{LLM Multi-Agent Systems: Challenges and Open Problems}

\icmlsetsymbol{equal}{*}

\begin{icmlauthorlist}
\icmlauthor{Shanshan Han}{uci}
\icmlauthor{Qifan Zhang}{uci}
\icmlauthor{Weizhao Jin}{usc}
\icmlauthor{Zhaozhuo Xu}{stevens}

\end{icmlauthorlist}

\icmlaffiliation{uci}{University of California, Irvine, CA, USA}
\icmlaffiliation{usc}{University of Southern California, Los Angeles, CA, USA}
\icmlaffiliation{stevens}{Stevens Institute of Technology, Hoboken, NJ, USA}

\icmlcorrespondingauthor{Shanshan Han}{shanshan.han@uci.edu}

\icmlkeywords{Machine Learning, ICML}

\vskip 0.3in
]

\printAffiliationsAndNotice{} 

\begin{abstract}

This paper explores multi-agent systems and identify challenges that remain inadequately addressed. By leveraging the diverse capabilities and roles of individual agents, multi-agent systems can tackle complex tasks through agent collaboration. We discuss optimizing task allocation, fostering robust reasoning through iterative debates, managing complex and layered context information, and enhancing memory management to support the intricate interactions within multi-agent systems. We also explore potential applications of multi-agent systems in blockchain systems to shed light on their future development and application in real-world distributed systems.

\end{abstract}

\section{Introduction}

Multi-agent systems enhance the capabilities of single LLM agents by leveraging collaborations among agents and their specialized abilities
~\cite{talebirad2023multi,zhang2023exploring,park2023generative,li2023camel,jinxin2023cgmi}. It utilizing collaboration and coordination among agents to execute tasks that are beyond the capability of any individual agent. 
In multi-agent systems, each agent is equipped with distinctive capabilities and roles, collaborating towards the fulfillment of some common objectives. Such collaboration, characterized by activities such as debate and reflection, has proven particularly effective for tasks requiring deep thought and innovation. 
Recent works include simulating interactive environments~\cite{park2023generative,jinxin2023cgmi}, role-playing~\cite{li2023camel}, reasoning~\cite{du2023debate,liang2023encouraging}, demonstrating the huge potential of multi-agent systems in handling complex real-world scenarios.

While existing works have demonstrated the impressive capabilities of multi-agent systems, the potential for advanced multi-agent systems far exceeds the progress made to date. A large number of existing works focus on devising planning strategies within a single agent by breaking down the tasks into smaller, more manageable tasks~\cite{chen2022program,ziqi-lu-2023-tab,yao2023tree,long2023large,besta2023graph,wang2022rationale}. 
Yet, multi-agent systems involve agents of various specializations and more complex interactions and layered context information, which poses challenges to the designing of the work flow as well as the whole system. 
Also, existing literature pays limited attention to memory storage, while memory plays a critical role in collaborations between agents. It enables agents to access to some common sense, aligning context with their tasks, 
and further, learn from past work flows and adapt their strategies accordingly. 

To date, multiple significant challenges that differentiate multi-agent systems and single-agent systems remain inadequately addressed. We summarize them as follows.

\begin{itemize}[leftmargin=2em, itemsep=-0.3em, topsep=0.3em]
    \item Optimizing task allocation to leverage agents' unique skills and specializations.
    \item Fostering robust reasoning through iterative debates or discussions among a subset of agents to enhance intermediate results.
    \item Managing complex and layered context information, such as context for overall tasks, single agents, and some common knowledge between agents, while ensuring alignment to the general objective.
    \item Managing various types of memory that serve for different objectives in coherent to the interactions in multi-agent systems 
\end{itemize}

This paper explores multi-agent systems, offering a survey of the existing works while shedding light on the challenges and open problems in it. We study major components in multi-agent systems, including planning and memory storage, and address unique challenges posed by multi-agent systems, compared with single-agent systems. We also explore potential application of multi-agent systems in blockchain systems from two perspectives, including 1) utilizing multi-agent systems as tools, and 2) assigning an agent to each blockchain node to make it represent the user, such that the agent can can complete some tasks on behalf of the user in the blockchain network.

\section{Overview}

\subsection{Structure of Multi-agent Systems}\label{sec: multi-agent-structure}

The structure of multi-agent systems can be categorized into various types, based on the each agent's functionality and their interactions. 
\begin{figure*}
    \centering
    \includegraphics[width=0.8\textwidth]{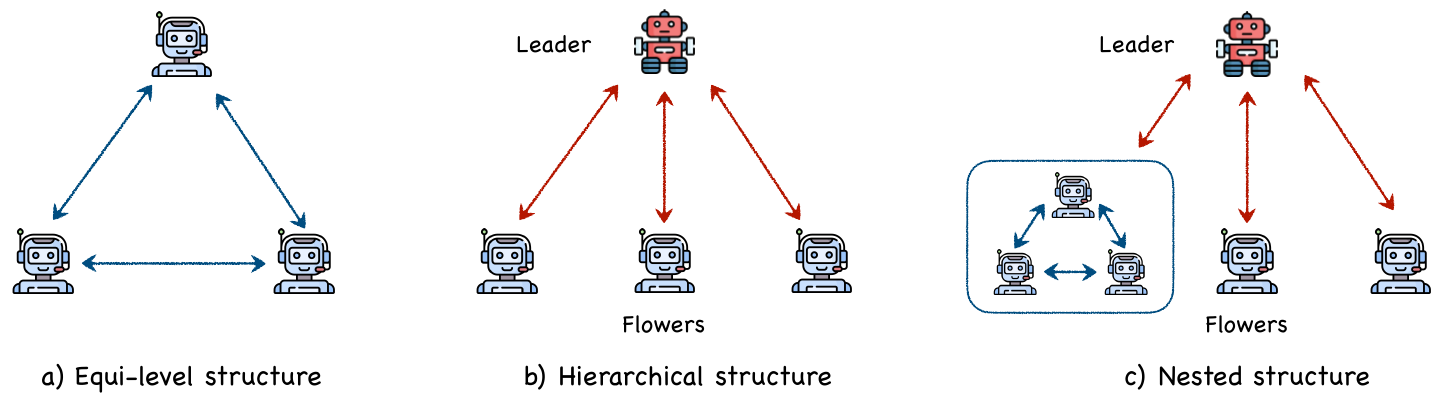}
    \caption{Structures of multi-agent systems.}
    \label{fig:agent_structure}
\end{figure*}

\textbf{Equi-Level Structure.}
LLM agents in an equi-level system operate at the same hierarchical level, 
where each agent has its role and strategy, but neither holds a hierarchical advantage over the other, \textit{e}.\textit{g}., DMAS~\cite{chen2023scalable}; see \cref{fig:agent_structure}(a). The agents in such systems can have same, neutral, or opposing objectives. 
Agents with same goals collaborate towards a common goal without a centralized leadership. The emphasis is on collective decision-making and shared responsibilities~\cite{li2019multi}. 
With opposing objectives, the agents negotiate or debate to convince the others or achieve some final solutions~\cite{terekhov2023second,du2023debate,liang2023encouraging,chan2023chateval}.  

\textbf{Hierarchical Structure.}
Hierarchical structures~\cite{survey_multi_agent_deep_learning,ahilan2019feudal} typically consists of a leader and one or multiple followers; see \cref{fig:agent_structure}(b). The leader's role is to guide or plan, while the followers respond or execute based on the leader's instructions. Hierarchical structures are prevalent in scenarios where coordinated efforts directed by a central authority are essential. 
Multi-agent systems that explore Stackelberg games~\citep{von2010market,conitzer2006computing} fall into this category~\citep{harris2023stackelberg}. This type of game is distinguished by this leadership-followership dynamic and the sequential nature of decision-making. 
Agents make decisions in a sequential order, where 
the leader player first generate an output (\textit{e}.\textit{g}., instructions) then the follower players take an
action based on the leader's instruction.

\textbf{Nested Structure.}
Nested structures, or hybrid structures, constitute sub-structures of equi-level structures and/or hierarchical structures in a same multi-agent system~\cite{chan2023chateval}; see \cref{fig:agent_structure}(c).
The ``big picture'' of the system can be either equi-level or hierarchical, however, as some agents have to handle complex tasks, they  break down the tasks into small ones and construct a sub-system, either equi-level or hierarchical, and ``invite'' several agents to help with those tasks. 
In such systems, the interplay between different levels of hierarchy and peer-to-peer interaction contributes to complexity. Also, the interaction among those different structures can lead to intricate dynamics, where strategies and responses become complicated due to the presence of various influencing factors, including external elements like context or environment.

\textbf{Dynamic Structure.}
Dynamic structures mean that the states of the multi-agent system, \textit{e}.\textit{g}., the role of agents, their relations, and the number of agents in the multi-agent system, may change~\cite{talebirad2023multi} over time. As an example, ~\cite{talebirad2023multi} enables addition and removal of agents to make the system to suit the tasks at hand. 
A multi-agent system may also be contextually adaptive, with the interaction patterns inside the system being modified based on internal system states or external factors, such as contexts. Agents in such systems can dynamically reconfigure their roles and relationships in response to changing conditions.

\subsection{Overview of Challenges in Multi-Agent Systems}
This paper surveys various components of multi-agent systems and discusses the challenges compared with single-agent systems. We discuss planning, memory management, as well as potential applications of multi-agent systems on distributed systems, \textit{e}.\textit{g}., blockchain systems.

\textbf{Planning.  }
In a single-agent system, planning involves the LLM agent  breaking down large tasks into a sequence of small, manageable tasks to achieve specific goals efficiently while enhancing interpretability, controllability, and flexibility~\cite{li2024personal,zhang2023igniting,nye2021show,wei2022chain}.
The agent can also learn to call external APIs for extra information that is missing from the model weights (often hard to change after pre-training), or connect LLMs with websites, software, and tools~\cite{patil2023gorilla,zhou2023webarena,cai2023large} to help reasoning and improve performance. While agents in a multi-agent system have same capabilities with single-agent systems, they encounter challenges inherited from the work flow in multi-agent systems. In \S\ref{sec: planning}, we discuss partitioning work flow and allocating the sub-tasks to agents; we name this process as ``global planning''; see \S\ref{sec: global_planning}. We then discuss task decomposition in each single-agent. Different from planning in a single-agent systems, agents in multi-agent systems must deal with more sophisticated contexts to reach alignment inside the multi-agent system, and further, achieve consistency towards the overall objective; see \S\ref{sec: single_agent_planning}. 

\textbf{Memory management. }
Memory management in single-agent systems include short-term memory during a conversation, long-term memory that store historical conversations, and, if any, external data storage that serves as a complementary information source for inferences, \textit{e}.\textit{g}., RAG~\cite{rag}. Memory management in multi-agent systems must handle complex context data and sophisticated interaction and history information, thus requires advanced design for memories. We classify the memories involved in multi-agent systems in \S\ref{sec: memory_classification} and then discuss potential challenges posed by the sophisticated structure of memory in \S\ref{sec: memory_challenges}.

\textbf{Application. }
We discuss applications of multi-agent systems in blockchain, a distributed system that involves sophisticated design of layers and applications. Basically, multi-agent systems can serve as a tool due to its ability to handle sophisticated tasks in blockchain; see \S\ref{sec: mas_tool_in_blockchain}. Blockchain can also be integrated with multi-agent systems due to their distributed nature, where an intelligent agent can be allocated to an blockchain node to perform sophisticated actions, such as negotiations, on  behalf of the agent; see \S\ref{sec: blockchain_nodes_as_agents}.

\section{Planning}\label{sec: planning}


Planning in multi-agent systems involves understanding the overall tasks and design work flow among agents based on their roles and specializations, (\textit{i}.\textit{e}., global planning) and breaking down the tasks for each agent into small manageable tasks (\textit{i}.\textit{e}., local planning). Such process must account for functionalities of the agents, dynamic interactions among the agents, as well as a more complex context compared with single-agent systems. This complexity introduces unique challenges and opportunities in the multi-agent systems.

\subsection{Global Planning}\label{sec: global_planning}

Global planning refers to understanding the overall task and split the task into smaller ones and coordinate the sub-tasks to the agents. It requires careful consideration of task decomposition and agent coordination. Below we discuss the unique challenges in global planning in multi-agent systems.

\textbf{Designing effective work flow based on the
agents’ specializations. }
Partitioning responsibilities and designing effective work flows for agents is crucial for ensuring that the tasks for each agent are executable while meaningful and directly contributes to the overall objective in systems. The biggest challenge lies in the following perspectives: 1) the partition of work flow should maximize the utilization of each agent's unique capabilities, \textit{i}.\textit{e}., each agent can handle a part of the task that matches its capabilities and expertise; 2) each agent's tasks must  align with the overall goal; and 3) the design must understand and consider the context for the overall tasks as well as each agent. This requires a deep understanding of the task at hand and the specific strengths and limitations of each agent in the system.

\textbf{Introducing loops for a subset of agents to enhance intermediate results. }
Multi-agent systems can be integrated with loops inside one or multiple subsets of agents to improve the quality of the intermediate results, or, local optimal answers. In such loops, agents debate or discuss to achieve an optimal results that are accepted by the agents in the loop. 
The iterative process can refine the intermediate results, leading to a deeper exploration of the task. The agents in the loop can adjust their reasoning process and plans during the loop, thus have better capabilities in handling uncertainties of the task.

\textbf{Game Theory. }
Game theory provides a well-structured framework for understanding strategic interactions in multi-agent systems, particularly for systems that involve complex interactions among agents such as debates or discussions. 
A crucial concept in game theory is equilibrium, \textit{e}.\textit{g}., Nash Equilibrium~\cite{kreps1989nash} and Stackelberg Equilibrium~\citep{von2010market,conitzer2006computing}, that  describes a state where, given the strategies of others, no agent benefits from unilaterally changing their strategy. Game theory has been applied in multi-agent systems, especially Stackelberg equilibrium~\cite{gerstgrasser2023oracles,harris2023stackelberg}, as the structure of
Stackelberg equilibrium contains
is a leader agent and multiple follower agents, and such hierarchical architectures are wildely considered in multi-agent systems.
\citep{gerstgrasser2023oracles}
designs a general multi-agent framework to identify Stackelberg Equilibrium in Markov games, and \cite{harris2023stackelberg} extend the Stackelberg model to allow agents to consider external context information, such as traffic and weather, etc.
However, some problems are still challenging in multi-agent systems, such as defining an appropriate payoff structure for both the collective strategy and individual agents based on the context of the overall tasks, and efficiently achieving equilibrium states. These unresolved issues highlight the ongoing need for refinement in the application of game theory to complex multi-agent scenarios.

\subsection{Single-Agent Task Decomposition}\label{sec: single_agent_planning}

Task decomposition in a single agent involves generating a series of intermediate reasoning steps to complete a task or arrive at an answer. This process can be represented as transforming direct input-output ($\langle \text{input}\rightarrow \text{output}\rangle$) mappings into the $\langle \text{input}\rightarrow\text{rational}\rightarrow\text{output}\rangle$ mappings~\cite{wei2022chain,zhang2023igniting}. Task composition can be of different formats, as follows.

\textit{i) Chain of Thoughts (CoT)}~\cite{wei2022chain} that transforms big tasks into step-by-step manageable tasks to represent interpretation of the agents' reasoning (or thinking) process.

\textit{ii) Multiple CoTs}~\cite{wang2022self} that explores multiple independent CoT reasoning paths  and return the one with the best output.


\textit{iii) Program-of-Thoughts (PoT)}~\cite{chen2022program} that uses language models to generate text and programming language statements, and finally an answer.

\textit{iv) Table-of-Thoughts (Tab-CoT)}~\cite{ziqi-lu-2023-tab}  that utilize a tabular-format for reasoning, enabling the complex reasoning process to be explicitly modelled in a highly structured manner.

\textit{v) Tree-of-Thoughts (ToT)}~\cite{yao2023tree,long2023large} that extends CoT by formulating a tree structure to explore multiple reasoning possibilities at each step. It enables generating new thoughts based on a given arbitrary thought and possibly backtracking from it. 

\textit{vi) Graph-of-Thoughts-Rationale (GoT-Rationale)}~\cite{besta2023graph} that explores an arbitrary graph to enable aggregating arbitrary thoughts into a new one and enhancing the thoughts using loops.

\textit{vii) Rationale-Augmented Ensembles}~\cite{wang2022rationale} that automatically aggregate across diverse rationales to overcome the brittleness of performance to sub-optimal rationales.

In multi-agent systems, task decomposition for a single agent becomes more intricate. Each agent must understand layered and sophisticated context, including 1) the overall tasks, 2) the specific context of the agent's individual tasks, and 3) the contextual information provided by other agents in the multi-agent system. 
Moreover, the agents must align these complex, multi-dimensional contexts into their decomposed tasks to ensure coherent and effective functioning within the overall task. 
We summarize the challenges for single agent planning as follows.

\textbf{Aligning Overall Context. }
Alignment of goals among different agents is crucial in multi-agent systems. Each LLM agent must have a clear understanding of its role and how it fits into the overall task, such that the agents can perform their functions effectively. Beyond individual roles, agents need to recognize how their tasks fit into the bigger picture, such that their outputs can harmonize with the outputs of other agents, and, further, ensuring all efforts are directed towards the common goal.

\textbf{Aligning Context Between Agents. }
Agents in multi-agent systems process tasks collectively, and each agent must understand and integrate the contextual information provided by other agents within the system to  ensure that the information provided by other agents is fully utilized.

\textbf{Aligning Context for Decomposed Tasks. }
When tasks of each agents are broken down into smaller, more manageable sub-tasks, aligning the complex context in multi-agent systems becomes challenging.  Each agent's decomposed task must fit their individual tasks and the overall goal while integrating with contexts of other agents. 
Agents must adapt and update their understanding of the task in response to context provided by other agents, and further, plan the decomposed tasks accordingly.

\textbf{Consistency in Objectives. }
In multi-agent systems, consistency in objectives is maintained across various levels, \textit{i}.\textit{e}., from overall goals down to individual agent tasks and their decomposed tasks. Each agent must understand and effectively utilize the layered contexts while ensuring its task and the decomposed sub-tasks to remain aligned with the overall goals. 
\cite{harris2023stackelberg} extends the Stackelberg model~\citep{von2010market,conitzer2006computing} to enable agents to incorporate external context information, such as context (or insights) provided by other agents. However, aligning the complex context with the decomposed tasks during reasoning remains unresolved.

\section{Agent Memory and Information Retrieval}\label{sec: memory}
The memory in single-LLM agent systems refers to the agent's ability to record, manage, and utilize data, such as past historical queries and some external data sources, to help inference and enhance decision-making and reasoning~\citep{yao2023tree,park2023generative,li2023mot,wang2023augmenting,guo2023prompt}.
While the memory in a single-LLM agent system primarily focuses on internal data management and utilization, a multi-agent system requires agents to work collaboratively to complete some tasks, necessitating the individual memory capabilities of each agent as well as a sophisticated mechanism for sharing, integrating, and managing information across agents, thus poses challenges to memory and information retrieval.

\subsection{Classifications of Memory in Multi-agent Systems}\label{sec: memory_classification}
Based on the work flow of a multi-agent system, we categorize memory in multi-agent system as follows.

\begin{itemize}[leftmargin=2em, itemsep=-0.3em, topsep=0em]
    \item \textit{Short-term memory:} This is the immediate, transient memory used by a Large Language Model (LLM) during a conversation or interaction, \textit{e}.\textit{g}., working memory in ~\cite{jinxin2023cgmi}. It is ephemeral, existing only for the duration of the ongoing interaction and does not persist once the conversation ends.
    \item \textit{Long-term Memory}: This type of memory stores historical queries and responses, essentially chat histories from earlier sessions, to support inferences for future interactions. Typically, this memory is stored in external data storage, such as a vector database, to facilitate recall of past interactions.

    \item \textit{External data storage}: This is an emerging area in LLM research where models are integrated with external data storage like vector databases, such that the agents can access additional knowledge from these databases, enhancing their ability to ground and enrich their responses~\cite{rag}. This allows the LLM to produce responses that are more informative, accurate, and highly relevant to the specific context of the query. 

    \item \textit{Episodic Memory}: This type of memory encompasses a collection of interactions within multi-agent systems. It plays a crucial role when agents are confronted with new tasks or queries. By referencing past interactions that have contextual similarities to the current query, agents can significantly enhance the relevance and accuracy of their responses. Episodic Memory allows for a more informed approach to reasoning and problem-solving, enabling a more adaptive and intelligent response mechanism, thus serves as a valuable asset in the multi-agent system, 

    \item \textit{Consensus Memory}: In a multi-agent system where agents work on a task collaboratively, consensus memory acts as a unified source of shared information, such as  common sense, some domain-specific knowledge, etc, \textit{e}.\textit{g}., skill library in~\cite{jinxin2023cgmi}. Agents utilize consensus memory to align their understanding and strategies with the tasks, thus enhancing an effective and cohesive collaboration among agents.
\end{itemize}

While both single-agent and multi-agent systems handle short-term memory and long-term memory, multi-agent systems introduce additional complexities due to the need for inter-agent communication, information sharing, and adaptive memory management.

\subsection{Challenges in Multi-agent Memory Management}\label{sec: memory_challenges}
Managing memory in multi-agent systems is fraught with challenges and open problems, especially in the realms of safety, security, and privacy. We outline these as follows:

\textbf{Hierarchical Memory Storage:} In a multi-agent system, different agents often have varied functionalities and access needs. Some agents may have to query their sensitive data, but they don't want such data to be accessed by other parties.
While ensuring the consensus memory to be accessible to all clients, implementing robust access control mechanisms is crucial to ensure sensitive information of an agent is not accessible to all agents. Additionally, as the agents in a system collaborative on one task, and their functionalities share same contexts, their external data storage and memories may overlap. If the data and functionalities of these agents are not sensitive, adopting an unified data storage  can effectively manage redundancy among the data, and furthermore, ensure consistency across the multi-agent system, leading to more efficient and precise maintenance of memory.

\textbf{Maintenance of Consensus Memory:} As consensus memory is obtained by all agents when collaborating on a task,
ensuring the integrity of shared knowledge is critical to ensure the correct execution of the tasks in the multi-agent systems. 
Any tampering or unauthorized modification of consensus memory can lead to systemic failures of the execution. 
Thus, a rigorous access control is important to mitigate risks of data breaches.

\textbf{Communication and information exchange:}
Ensuring effective communication and information exchange between agents is essential in multi-agent systems. Each agent may hold critical pieces of information, and seamless integration of these is vital for the overall system performance. 

\textbf{Management of Episodic Memory. } Leveraging past interactions within the multi-agent system to enhance responses to new queries is challenging in multi-agent systems. Determining how to effectively recall and utilize contextually relevant past interactions among agents for current problem-solving scenarios is important.

These challenges underscore the need for continuous research and development in the field of multi-agent systems, focusing on creating robust, secure, and efficient memory management methodologies.

\section{Applications in Blockchain}\label{sec: blockchain_app}

Multi-agent systems offer significant advantages to blockchain systems by augmenting their capabilities and efficiency. 
Essentially, these multi-agent systems serve as sophisticated tools for various tasks on blockchain and Web3 systems. 
Also, 
blockchain nodes can be viewed as agents with specific roles and capabilities~\cite{ankile2023see}.
Given that both Blockchain systems and multi-agent systems are inherently distributed, the blockchain networks can be integrated with multi-agent systems seamlessly. 
By assigning a dedicated agent to each blockchain node, it's possible to enhance data analyzing and processing while bolstering security and privacy in the chain.

\subsection{Multi-Agent Systems As a Tool}\label{sec: mas_tool_in_blockchain}
To cast a brick to attract jade, we give some potential directions that multi-agents systems can act as tools to benefit blockchain systems.

\textbf{Smart Contract Analysis. }
Smart contracts are programs stored on a blockchain that run when predetermined conditions are met. 
Multi-agents work together to analyze and audit smart contracts. The agents can have different specializations, such as identifying security vulnerabilities, legal compliance, and optimizing contract efficiency. Their collaborative analysis can provide a more comprehensive review than a single agent could achieve alone.

\textbf{Consensus Mechanism Enhancement.} Consensus mechanisms like Proof of Work (PoW)~\cite{gervais2016security} or Proof of Stake (PoS)~\cite{saleh2021blockchain} are critical for validating transactions and maintaining network integrity. Multi-agent systems can collaborate to monitor network activities, analyze transaction patterns, and identify potential security threats. By working together, these agents can propose enhancements to the consensus mechanism, making the blockchain more secure and efficient.

\textbf{Fraud Detection. } 
Fraud detection is one of the most important task in financial monitoring. As an example, \cite{ankile2023see} studies fraud detection through the perspective of an external observer who detects price manipulation by analyzing the transaction sequences or the price movements of a specific asset.
Multi-agent systems can benefit fraud detection in blockchain as well. Agents can be deployed with different roles, such as monitoring transactions for fraudulent activities and analyzing user behaviors. Each agent could also focus on different behavior patterns to improve the accuracy and efficiency of the fraud detection process.

\subsection{Blockchain Nodes as Agents}\label{sec: blockchain_nodes_as_agents}
\cite{ankile2023see} identifies  blockchain nodes as agents, and studies fraud detection in the chain from the perspective an external observer. However, as powerful LLM agents with analyzing and reasoning capabilities, there are much that the agents can do, especially when combined with game theory and enable the agents to negotiate and debate. Below we provide some perspectives.

\textbf{Smart Contract Management and Optimization.}
Smart contracts are programs that execute the terms of a contract between a buyer and a seller in a blockchain system. The codes are fixed, and are self-executed when predetermined conditions are met. 
Multi-agent systems can automate and optimize the execution of smart contracts with more flexible terms and even dynamic external information from users. Agents can negotiate contract terms on behalf of their users, manage contract execution, and even optimize gas fees (in the context of Ethereum~\cite{wood2014ethereum}. The agents can analyze context information , such as past actions and pre-defined criteria, and utilize the information with flexibility.
Such negotiations can also utilize game theory, such as Stackelberg Equilibrium~\citep{von2010market,conitzer2006computing} when there is a leader negotiator and Nash Equilibrium~\cite{kreps1989nash} when no leader exists.

\section{Conclusion}

The exploration of multi-agent systems in this paper underscores their significant potential in advancing the capabilities of LLM agents beyond the confines of single-agent paradigms. By leveraging the specialized abilities and collaborative dynamics among agents, multi-agent systems can tackle complex tasks with enhanced efficiency and innovation. Our study has illuminated challenges that need to be addressed to harness the power of multi-agent systems better, including optimizing task planning, managing complex context information, and improving memory management. Furthermore, the potential applications of multi-agent systems in blockchain technologies reveal new avenues for development, which suggests a promising future for these systems in distributed computing environments. 


\bibliography{example_paper}
\bibliographystyle{icml2024}

\newpage
\appendix

\end{document}